\documentclass[12pt]{article}

\usepackage{amsmath,amssymb}
\usepackage{graphicx}
\usepackage[dvips]{color} 

\makeatletter

\@addtoreset{equation}{section}
\makeatother

\newcommand{\be}{\begin{equation}}
\newcommand{\ee}{\end{equation}}
\newcommand{\bea}{\begin{eqnarray}}
\newcommand{\eea}{\end{eqnarray}}
\newcommand{\beann}{\begin{eqnarray*}}
\newcommand{\eeann}{\end{eqnarray*}}

\newcommand{\ba}{\begin{array}}
\newcommand{\ea}{\end{array}}
\newcommand{\Slash}[1]{{\ooalign{\hfil$#1$\hfil\crcr\raise.167ex\hbox{/}}}}
\newcommand{\SL}{\Slash{L}}

\newcommand{\Tr}{\mathop{\rm Tr}}

\newcommand{\N}{{\cal N}}

\def\XXint#1#2#3{{\setbox0=\hbox{$#1{#2#3}{\int}$} 
\vcenter{\hbox{$#2#3$}}\kern-.5\wd0}}

\begin{document}

\setlength{\oddsidemargin}{0cm}
\setlength{\baselineskip}{7mm}

\begin{titlepage}
\renewcommand{\thefootnote}{\fnsymbol{footnote}}


\begin{flushright}
\begin{tabular}{l}
RIKEN-TH-158\\
KUNS-2210
\end{tabular}
\end{flushright}

~~\\
\vspace*{0cm}
    \begin{Large}
    \begin{bf}
       \begin{center}
         { Eschenburg space as gravity dual  \\
of \\
flavored ${\mathcal N}$=4 Chern-Simons-matter theory
} 
      
       \end{center}
    \end{bf}   
    \end{Large}
\vspace{1cm}

\begin{large}
\begin{center}
{ \sc Mitsutoshi Fujita}\footnote    
{e-mail address : 
mfujita@gauge.scphys.kyoto-u.ac.jp} and 
{ \sc Ta-Sheng Tai}\footnote    
{e-mail address : 
tasheng@riken.jp} 
\end{center}
\end{large}

      \vspace{1cm}

\begin{center}{\it Department of Physics, Kyoto University, Kyoto 606-8502, JAPAN}\\
{\it   Theoretical Physics Laboratory, RIKEN,
                    Wako, Saitama 351-0198, JAPAN}
                    
\end{center}

\vspace{1cm}

\begin{abstract}
\noindent
We find a 3D 
flavored ${\mathcal N}$=4 Chern-Simons-matter theory, 
a kind of ${\mathcal N}$=3 SCFT, 
has a gravity dual 
$AdS_4\times {{\mathcal M}_7 (t_1,t_2,t_3)}$ where 
three coprime 
parameters can be read off according to the number and charge of 5-branes in the dual Type IIB string setup. 
Because ${{\mathcal M}_7 (t_1,t_2,t_3)}$ has been known 
in literature as 
Eschenburg space, we exploit some of its properties 
to examine the correspondence between two sides. 

\end{abstract}
\vfill 

\end{titlepage}
\vfil\eject

\setcounter{footnote}{0}

\section{Introduction}The program towards studying gauge/gravity correspondence in the context of AdS$_4$/CFT$_3$ \cite{Maldacena:1997re} becomes concrete owing 
to the pioneering work \cite{Aharony:2008ug} by Aharony, Bergman, Jafferis 
and Maldacena last year. They found that constructing a much higher supersymmetric conformal field theory (SCFT) of Chern-Simons-matter (CSM) type is possible due to an elliptic brane setup in Type IIB string theory. Through T-duality and 
M-theory lift, one obtains $N$ M2-branes filling (012) transverse to a 8D cone: Cone$(\mathcal{B}_7)$ 
$({\mathcal B}_7=S^7/{\bf Z}_k)$
 along (345678910). 
The corresponding gravity dual is thus 
a solution of 11D supergravity 
$AdS_4\times {\mathcal B}_7$ 
after $N$ M2-branes backreact.

Later on, generalizing their idea to yield elliptic 
${\mathcal N}\ge3$ SCFTs is explained in 
\cite{Imamura:2008nn, Jafferis:2008qz} by attaching various kinds of 
$(1,k_i)$5-branes 
on a stack of circular D3-branes. The resulting field theory at 
infra-red (IR) fixed point $(g_{YM}\to\infty)$ 
is still of quiver CSM type with product 
gauge group $\prod_I U(N)_I$. Its Lagrangian is so 
rigid, i.e. CS level $k_I$ w.r.t. $I$-th gauge factor is  determined by two adjacent 
D5-brane charges $ k_I=k_i-k_{i-1}$ \cite{Ohta, Ber}, while the 
superpotential is obtained by integrating out non-dynamical massive adjoint 
$\Phi_I (\subset$ vector multiplet) coupled to 
hypermultiplets in a typical manner. 

Among many kinds of elliptic 
${\mathcal{N}}\ge3$ SCFTs, we will focus on a specific type of ${\mathcal N}$=4 SCFT which is constructed via IIB $N$ circular D3- (0126), 
$p$ NS5- (012345) and $q$ $(1,k)$5- 
($012[3,7]_{\theta}[4,8]_{\theta}[5,9]_{\theta}$) branes%
\footnote 
{$\theta$ (twisted angle) and 
${g^2_{YM}k}/{4\pi}$ (adjoint mass) are related to each other by ($L$: segment length on $x^6$)  
\begin{eqnarray*}
\frac{\tan \theta}{L}=g^2_{YM} k ,
~~~~~\frac{1}{g^2_{YM}}=\frac{L}{g_s}.
\label{fract}
\end{eqnarray*}
Taking IR limit implies naturally 
a strongly coupled M-theory picture. }. 
Its 11D gravity dual 
$AdS_4\times {\mathcal M}_7$ parameterized by $(k,p,q)$ 
is explicitly known \cite{Imamura:2008nn, Imamura:2008ji} and this is the main 
reason why we study this kind of SCFT here.

In this note, motivated by works on 
the flavored ABJM theory \cite{Hohenegger:2009as,  Gaiotto:2009tk, Hikida:2009tp}, 
we construct a new ${\mathcal N}$=3 SCFT 
by adding $N_F$ massless fundamental flavors and study its gravity dual. From Type IIB picture, adding flavor corresponds to further attaching $N_F$ D5-branes (012789) on the circle $x^6$ and results in a less supersymmetric ${\mathcal N}$=3 SCFT. 
This construction is by definition an elliptic one, so 
in M-theory to have $N$ M2-branes probing a 8D 
cone may thus be expected due to conformality.

We find that this turns out to be true and 
the dual geometry is now 
$AdS_4\times \tilde{{\mathcal M}}_7$ parameterized by three natural 
numbers $(t_1, t_2, t_3)=(qN_F, pN_F,kpq)$ without any common factor. 
In fact, many properties of 
$\tilde{{\mathcal M}}_7(t_1, t_2, t_3)$ (modulo common factor) known as 
Eschenburg space \cite{E} 
have been 
explored by mathematicians \cite{1,2,3}. 
For example, Cone($\tilde{{\mathcal M}}_7$) is Ricci-flat 
with special $Sp(2)$ holonomy. Namely, it is 
hyperK{\"a}hler and the base $\tilde{{\mathcal M}}_7$ must be tri-Sasakian (which preserves a fraction
3/16 of 32 SUSY). 
Moreover, the cone is available through applying a 
hyperK{\"a}hler quotient to a 3D (flat) quaternionic 
space, say, 
${\bf H}^3///U(1)_Q \equiv \boldsymbol{\mu}_Q^{-1}(0)/U(1)_Q$\footnote{Certainly, one can solve D- and F-term 
conditions to see that M2-brane moduli space is 
${\bf {Sym}}^N ({\mathbb M})$ with  
${\mathbb M} \sim {\bf H}^{3}///{\text { Ker}} (\beta)$ 
and%

\begin{eqnarray}
\beta: U(1)^{3} \to U(1)^2, ~~~~~~~
\beta=\left(
\begin{array}{ccc}
p& q & 0\\
0 &  kq & N_F 
\end{array}
\right).
\label{111}
\end{eqnarray}
The equivalence between 
two descriptions seems straightforward because three 
moment maps $\boldsymbol{\mu}_Q=\sum^3_{i=1} t_i \boldsymbol{\mu}_i$, 
$p\boldsymbol{\mu}_1 + q\boldsymbol{\mu}_2$ 
and $kq\boldsymbol{\mu}_2 + N_F \boldsymbol{\mu}_3$ 
are linearly independent. }
 and $(t_1, t_2, t_3)$ stands for 
the underlying $U(1)_Q$ charge assignment respectively for three 
quaternions%
\footnote{Note that 
the amount of its isometry is essentially 
$SU(2)_R \times U(1)^2$ but $U(1)^2$ gets enhanced to 
$SU(2)\times U(1)$ ($SU(3)$) if two (all) of three $t$'s coincide.}.

In order to understand relations better 
between the three, say, 
flavored ${\mathcal N}$=4 CSM, $AdS_4 \times \tilde{\mathcal{M}}_7$ M-theory dual and 
IIA gravity dual of ${\mathcal N}$=4 CSM 
with probe flavor branes embedded, 
we adopt the viewpoint similar to \cite{Gubser:1996de,  Nishioka:2007zz}. 
That is, we compute the entropy 
of the three. 
On the field theory side, we take large $N$ zero-coupling 
limit and 
compactify $R^{1,2}$ on $S^1 \times S^2$. Its 
partition 
function is finally expressed in terms of an unitary matrix model which is 
exactly solvable. 
A similar formulation using a matrix model 
allows us to evaluate 
${\mathcal N}$=3 superconformal index to which 
we are able to compare Kaluza-Klein (KK) analysis on 
$\tilde{{\mathcal M}}_7$.

On the geometry side, we are led to 
compute the volume of 7-cycle of 
Eschenburg space by taking advantage of a formula given in \cite{Lee:2006ys}. 
Also, the on-shell action 
of IIA probe D6-branes is taken care of. 
As pointed out in \cite{Gaiotto:2009tk}, 
the correct embedding of D6-branes can be found 
by performing a further 
hyperK{\"a}hler quotient to 
Cone(${{\mathcal M}}_7$). 
One obtains a 4D Taub-NUT space thereof 
over which flavor probes  
should wrap after doing KK reduction to IIA theory. 
In addition, we consider 5-cycles among 
$\tilde{{\mathcal M}}_7$ because 
M5-branes wrapped over them correspond to 
baryonic operators in the field theory.

This note is organized as follows. 
We begin with 
discussing how to obtain a 3D ${\mathcal N}$=3 SCFT from 
an elliptic ${\mathcal N}$=4 one. Then 
in section 3 and section 4, we study the entropy 
of our underlying ${\mathcal N}$=3 SCFT by means of 
both field theory and gravity approaches. 
An ${\mathcal N}$=3 superconformal index is computed in 
section 5 and comments about 
baryonic operators are in section 6. Finally, 
a conclusion is drawn. Appendices about Taub-NUT 
space in M-theory and mesonic 
operators are attached.

\section{Adding flavors to 3D ${\mathcal N}$=4 SCFT}
In section 2.1, we shortly review some aspects 
about $\mathcal{N}$=4 Chern-Simons-matter theory \cite{Imamura:2008nn, H, I}. 
In section 2.2, by adding massless flavors to it, a new $\mathcal{N}$=3 SCFT is constructed. 
We observe that this $\mathcal{N}$=3 Lagrangian 
requires naturally a 8D hyperK{\"a}hler internal space. 

\subsection{${\mathcal N}$=4 SCFT} 
In order to obtain a desired ${\mathcal N}$=4 SCFT at IR fixed point, 
one begins with an 
ultra-violet (UV) Lagrangian ${\mathcal L}^{UV}$ containing ${\mathcal N}$=4 $(V^I, \Phi^I)$ vector-, 
$(A^I, B^I)$ hyper-, and 
$(A^J, B^J)$ twisted hyper-multiplets where $I$ labels 
the gauge 
factor. ${\mathcal L}^{UV}$ can be read off from the corresponding 
IIB brane 
configuration (or quiver diagram) considered in section 1%
\footnote{In fact, there are some thoughts in dealing with 
$p\ne q$ as shown in \cite{I} for zero CS levels, but we will ignore these subtleties. }. 
Due to CS terms induced, vector multiplets acquire mass 
$\sim k_I g^2_{YM}/ 4\pi$ and at 
low-energy limit 
($g_{YM}\to\infty$) 
kinetic terms of 
them proportional to $1/ g^{2}_{YM}$ all decouple except for CS terms which 
do not depend on $g_{YM}$. 
The remaining non-dynamical 
adjoint $\Phi^I$ in F-terms 
or real scalar $\sigma^I \subset V^I$ in D-terms 
will be integrated out later. As a result, one arrives 
at a 3D ${\mathcal N}$=4 CSM theory with 
${\mathcal L}_{bos}^{IR} = {\mathcal L}_{CS} + {\mathcal L}_{hyper} + 
{\mathcal L}_{pot}$ where (bosonic part only) 
\begin{eqnarray*}
{\mathcal L}_{hyper} = \Tr \sum_I 
\int d^3 x  d^4 \theta ~ (\overline{A}^{ I }e^{2V^I} A^I e^{-2V^{I+1}}
+ {B}^I e^{-2V^I} \overline{B}^{ I} e^{2V^{I+1}}), \nonumber\\
{\mathcal L}_{pot} =\Tr \sum_I \frac{1}{k_I} \int d^3 x d^2 \theta (B^I A^I - A^{I-1} B^{I-1})^2 + c.c. .
\end{eqnarray*} 
Unlike ${\mathcal N}$=3 Chern-Simons-Yang-Mills (CSYM) 
theory obtained by 
adding CS terms to ${\mathcal N}$=4 YM one \cite{Kao:1992ig, Kao:1993gs},  
that YM terms decouple here, on the contrary, 
doubles the amount of SUSY. The 
$SO(4)_R$ $R$-symmetry arises from 
$SU(2)_{t} \times SU(2)_{unt}$ rotating 
$(A^I, B^I)$ 
and $(A^J, B^J)$ which are 
massless open string modes across $(1,k)$5- and NS5-branes, respectively%
\footnote{Note that $A^I \supset (h^I_\alpha, \psi^I_{\dot{\alpha}})$, 
$B^I \supset (\tilde{h}^I_\alpha, \tilde{\psi}^I_{\dot{\alpha}
})$ and 
$A^J\supset (h^J_{\dot{\alpha}}, \psi^J_{{\alpha}})$, 
$B^J \supset (\tilde{h}^J_{\dot{\alpha}}, 
\tilde{\psi}^J_{{\alpha}})$ where $\alpha$ ($\dot{\alpha}$) denotes 
the spinor index of $SU(2)$.}. With 
the baryonic $U(1)_b$ and diagonal $U(1)_d$, 
these as a whole agree precisely with the isometry $\big( SU(2)\times U(1)\big)^2$ of its 
moduli space 
$\big( {\bf C}^2/{\bf Z}_p\times {\bf C}^2/{\bf Z}_{q} \big)/{\bf Z}_k$ \cite{Imamura:2008nn}. 
The global symmetry of ${\mathcal N}$=4 CSM theory is summarized in Table \ref{1}. 
\begin{table}[h]
\caption{The global symmetry of $\mathcal{N}=4$ CSM theory}
  \label{1}
  \begin{center}
    \begin{tabular}{c|c|c|c|c|c|c} \hline
&$U(N)_{I-1}$&$U(N)_{I}$&$SU(2)_R $& $SU(2)_R$&$U(1)_b$&$U(1)_d$\\ \hline
 $h^I_{\alpha}$  &${\bf N}$&$\overline{{\bf N}}$&${\bf 2}$&${\bf 1}$&1&1\\ \hline
  $h^I_{\dot{\alpha}}$  &${\bf N}$&$\overline{{\bf N}}$&${\bf 1}$&${\bf 2}$&-1&1\\ \hline
    \end{tabular}
  \end{center}
\end{table}

In summary, 
in IR limit the action consists of only CS terms, kinetic 
terms of hyper-multiplets and a suitable superpotential. 
A comment is as follows. The structure of Lagrangian is quiet simple and one can think 
that CS terms impose a special 
kind of $gauging$ of $(n-2)$ out of $n$ $U(1)$ factors%
\footnote{Assume $N$=1 and $n$: \# of 5-branes. }, 
except for an overall diagonal $U(1)$ and the 
dual photon ${\mathcal{A}_{\mu}} = \sum_I k_I A^I_{\mu}$. Note that the 
remnant of ${\mathcal{A}_{\mu}}$ is some discrete gauge symmetry.

Due to the $(n-2)$ $gauging$, a direct observation is that the 
maximal global symmetry $SO(8)$ of moduli space is broken to 
${\mathcal{G}} \times U(1)^2$, 
where $U(1)^2$ correspond to ungauged ones. 
To determine ${\mathcal{G}}$ relies on knowing the 
complete moduli space.  
In ${\mathcal{N}}$=6 ABJM case, the diagonal $U(1)$ gets included in the $SU(4)_R$ $R$-symmetry. 
In ellptic ${\mathcal{N}}$=4 models above, ${\mathcal{G}} = SO(4)_R$ does not get mixed with 
two $U(1)$'s.

\subsection{Adding flavors }
Next, let us add massless flavors to ${\mathcal N}=4$ SCFT, i.e. 
to $I$-th gauge group $N^I_F$ fundamental hypermultiplets
$(Q^I, \tilde{Q}^I)$ of 
$({\bf N},\overline{\bf{N}})$ with $\sum _I N^I_F =N_F$. 
This results in an additional D-term 
\begin{eqnarray*}
{\mathcal L}_{flavor} = \Tr \sum_{\alpha, I} 
\int d^3 x  d^4 \theta ~ (\overline{Q}_{\alpha}^{ I }e^{2V^I} 
Q_{\alpha}^I 
+ \tilde{Q}_{\alpha}^I e^{-2 V^I} \overline{\tilde{Q}}_{\alpha}^{ I} ), \quad\quad \alpha=1,\cdots,N_F^I 
\end{eqnarray*}
and 
\begin{eqnarray*}
{\mathcal L}_{pot}\to{\mathcal L}'_{pot}
=\Tr \sum_{\alpha, I} \frac{1}{k_I} \int d^3 x d^2 \theta   (B^I A^I - A^{I-1} B^{I-1} +Q_{\alpha}^{I-1} \tilde{Q}_{\alpha}^{I-1}  )^2 + c.c..
\end{eqnarray*}
The $R$-symmetry is now broken to $SO(3)_R$ 
which is the diagonal 
$SU(2)_d \subset SU(2)_{R} \times SU(2)_{R} \simeq SO(4)_R $, 
while $U(1)_b \times U(1)_d$ 
stays unchanged. 
We find this is consistent with the amount of isometry of 
Eschenburg space (with three different $t$'s) 
as advertised in footnote 2. 

\section{Entropy from field theory }

Let us do a very 
simple counting of the degrees of freedom in ${\mathcal N}$=3 SCFT. This is carried out by computing 
the entropy of a dilute gas of massless 
states via statistical mechanics. 
The system is put in a box of size $V_2=L^2$ and momenta of massless 
states are quantized as $\vec{p}=2\pi {\vec{n}}/{L}$ 
$(\vec{n}\in {\bf Z}^2)$. 
\begin{eqnarray*}\label{sfra}
S=-\frac{\partial F}{\partial T},\quad \quad 
F=-T\log Z = 
T \frac{V_2}{4\pi^2} \sum_{i=1,2} 
\int d^2p ~s_i \log (1-s_i e^{-\beta {\mathcal E}})
\end{eqnarray*}
where $s_i =\pm$ and 
${\mathcal E}=\sqrt{p_1^2 + p_2^2}$. 
Therefore, 
\begin{eqnarray*}\label{2pii}
2\pi \int d{\mathcal E}  
\log \big( \tanh (\frac{1}{2} \beta {\mathcal E})\big){\mathcal E}
=-\frac{2\pi }{\beta^2} \frac{7}{4} \zeta(3),\\ 
S=\frac{21}{2\pi} N^2 V_2 T^2 \zeta(3) (p+q+\frac{N_F}{N}) 
+ {\mathcal O}(\lambda).
\end{eqnarray*}
For $p=q=1$ and $N_F=0$, ABJM result in \cite{Aharony:2008ug} is reproduced. The power of $N$ here, namely, 
$N^2$ deviates from $N^{\frac{3}{2}}$ derived from the gravity result. This problem remains unsolved because we are just using 
the gauge theory on M2-branes.

\subsection{Matrix model free energy}
Let us try another method to compute 
the entropy (or free energy) of SCFTs 
in large $N$ limit with 't Hooft coupling $\lambda =N/k \ll 1$. 
We assume for simplicity 
NS5- and $(1,k)$5-branes are
placed pairwise on the circle $x^6$ such that 
$p=q$ in Type IIB setup. 
One needs to 
compute the unitary matrix integral ($x=e^{-\beta}$) \cite{S1, A1, NTP}: 
\begin{eqnarray}
Z=\int \prod^{2q}_{I=1} DU_I 
\exp  \sum^q_{i=1} \sum^{\infty}_{n=1} 
\frac{1}{n} \Big( z^{unt} (x^n) \big(  \Tr (U^n_{2i})  \Tr (U^{-n}_{2i+1}) + (n \leftrightarrow -n)  \big) \nonumber\\
+ z^{{t}} (x^n) \big( \Tr (U^n_{2i-1})  \Tr (U^{-n}_{2i}) + (n \leftrightarrow -n) \big) \nonumber \\
+ z_{2i}^f (x^n) \big(  \Tr (U^n_{2i}) + \Tr (U^{-n}_{2i}) \big) 
+ z_{2i-1}^f (x^n) \big( \Tr (U^n_{2i-1}) + \Tr (U^{-n}_{2i-1}) \big) \Big).
\label{ZZ}
\end{eqnarray} 
Note that $t$, $unt$ and $f$ stand for twisted, untwisted and 
flavor, respectively. 
The matrix model arises from compactifying CSM theory 
on $S^1 \times S^2$ $(t\sim t+\beta)$, taking suitable temporal gauge 
and integrating out matters. Here, Polyakov loop 
$U_I =e^{i\beta A_0^I}$ satisfies $U_{2q+1}=U_{1}$ and $U^{-1}=U^\dagger$. 

By writing the measure as $DU = \prod^{\infty}_{n=1} d\rho_n \exp (-N^2 \sum_n \dfrac{\rho_n \rho_{-n}}{n})$ with $\rho_n = \dfrac{1}{N} \Tr U^n$ (which facilitates large $N$ limit), 
$Z$ becomes 
\begin{eqnarray*}
Z=\int \prod_{i,n} d\rho_{i,n} d\chi_{i,n} \exp \sum_{i,n} -\frac{N^2}{n}
\Big(  \rho_{i,n}  \rho_{i,-n} +   \chi_{i,n}  
\chi_{i,-n}   \\
- \frac{1}{N} z^f_{2i,n} ( \rho_{i,n} +  \rho_{i,-n} ) 
- \frac{1}{N} z^f_{2i-1,n} ( \chi_{i,n} + \chi_{i,-n} ) \\
- z_n^{unt} \big(  \chi_{i,n} \rho_{i+1,-n} + (n \leftrightarrow -n) \big) 
-{z}_n^t \big( \rho_{i,n} \chi _{i,-n} + (n \leftrightarrow -n) \big)  \Big).
\end{eqnarray*} 
Notice that $\rho$ $(\chi)$ comes from the odd (even) subscript of $U_I$.  
Alternatively, in large $N$ limit, one can introduce 
an eigenvalue density function for each $U_I$ like 
\begin{eqnarray*}
\sigma_I  (\theta) =\dfrac{1}{2\pi} + 
\sum^{\infty}_{m=1}\dfrac{\rho_{I,m}}{\pi}
\cos(m\theta), \quad\quad \int_0^{2\pi} d\theta \sigma_I  (\theta) =1 
\end{eqnarray*} 
to solve the matrix model. 
Because 
$U$ appears only in characters of $U(N)$, 
one can just express in the diagonal form $U={diag}(e^{i\theta_1},\cdots,e^{i\theta_N})$ and get rid of 
irrelevant angular parts. 
Also, $DU$ is the invariant Haar measure normalized as $\int DU \Tr_{R'} U \cdot \Tr_R U^\dagger =\delta_{R' R}$. 
It is easily confirmed that either way leads to the same expression of $Z$.

Further taking high temperature limit $\beta \ll 1$ to 
facilitate the comparison with gravity results, 
we find that there is a saddle point 
$\rho_{i,n}=\chi_{i,n}=1$. 
Because of  
\begin{eqnarray*}
&z_n^t =z^{unt}_n =z_n =2\Big( 
z_B (x^n) + (-)^{n+1} z_F (x^n) \Big), 
\quad\quad 
z^f_{i,n}=N^i_F  z_n,\\
&z_B (x)=\dfrac{x^{\frac{1}{2}} \left( 1+x \right) }{ \left( 1-x  \right)^{2}}, \quad\quad
z_F (x) = \dfrac { 2x}{ \left( 1-x
\right) ^{2}},
\end{eqnarray*}
by using asymptotics of $z$'s 
\begin{eqnarray*}
&z_B(x^n)\to \dfrac{2}{(n\beta)^2}+
{\mathcal O}(\dfrac{1}{n\beta}), \quad\quad
&z_F(x^n)\to \dfrac{2}{(n\beta)^2}+{\mathcal O}(\dfrac{1}{n\beta})
\end{eqnarray*}
and $\zeta$-function, 
\begin{eqnarray*}
Z\sim \exp \sum^\infty_{n=1} \frac{2N^2}{n}\big(  \frac{N_F}{N} 
+ 2q    \big)z_n 
\end{eqnarray*}
gives rise to 
\begin{eqnarray*}
F=-2 T^3 \zeta(3)\frac{7}{4} {\mathcal N}, 
\quad\quad S=\frac{21}{2} 
T^2 \zeta(3) {\mathcal N},\quad\quad{\mathcal N}=4(NN_F + 2qN^2).
\end{eqnarray*}
We find agreement with the previous result up to some 
irrelevant 
constant $V_2 \pi /4\pi^2$.

\section{Entropy from gravity dual}
Now let us proceed to examine issues about the entropy (counting 
 degree of freedom) of the obtained $\mathcal{N}$=3 SCFT using both the 
11D M-theory dual and known IIA dual geometry with probe flavor branes 
embedded. The entropy obtained by making use of the free field theory approximation 
will therefore be compared with these 
gravity calculations.

To fulfill this purpose, we shall 
demonstrate more 
precisely the 8D transverse space to M2-branes 
is a 8D hyperK{\"a}hler manifold, 
Eschenburg space, whose isometry, holonomy, and volume 
have been known (see also Introduction).

\subsection{Eschenburg space as gravity dual}

According to the remarkable work of Gauntlett, Gibbons, Papadopoulos and Townsend \cite{GGPT}, one is able to 
have a dictionary translating certain IIB 5-brane configuration into a 11D M-theory geometry $R^{1,2} \times {\bf M}_8$. 
This works also in our case where ${\bf M}_8$ is now specified as Eschenburg space. As noted before, its 
hyperK{\"a}hler structure makes the 
symmetry match with 
the newly obtained $\mathcal{N}$=3 SCFT quite successful.

For $\varphi_i\in (0, 4\pi]$,  
\begin{eqnarray}
ds_{8D}^2 = \frac{1}{2} U_{ij} 
d {\boldsymbol{x}}_i \cdot  d {\boldsymbol{x}}_j + \frac{1}{2}U^{ij} (d\varphi_i + A_i)(d\varphi_j +A_j ),\nonumber\\
A_i=d{\boldsymbol{x}}_j \cdot \boldsymbol{\omega}_{ji}
=dx^a_j ~ \omega_{ji}^a, \quad \partial_{x^a_j}\omega^b_{ki}-\partial_{x^b_k} \omega_{ji}^a=\epsilon^{abc}\partial_{x^c_j} U_{ki}
\label{8}
\end{eqnarray}
where $i,j,k=1,2$ and $a,b,c=1,2,3$ (Cartesian label). Note that 
$U_{ij}$ is a 2 by 2 symmetric matrix:
\begin{eqnarray}
U_{ij}=\frac{1}{2}\left(
\begin{array}{cc}
\dfrac{p}{|{\boldsymbol{x}}_1| } +  
\dfrac{q}{|{\boldsymbol{x}}_1 + k{\boldsymbol{x}}_2|} & 
\dfrac{kq}{|{\boldsymbol{x}}_1 + k{\boldsymbol{x}}_2|}\\
\dfrac{kq}{|{\boldsymbol{x}}_1 + k{\boldsymbol{x}}_2|}&
\dfrac{k^2 q}{|{\boldsymbol{x}}_1 + k{\boldsymbol{x}}_2|}
\end{array}
\right)
\label{u}
\end{eqnarray}
in the case of $p$ NS5- and $q$ $(1,k)$5-branes on Type IIB side. Here, ${\boldsymbol{x}}_1=(345)$ and 
${\boldsymbol{x}}_2=(789)$. 
The normalization of $U$ is chosen such that 
it gives \eqref{Y7} after M2-brane backreaction. 

Let us perform the following $GL(2)$ transformation 
\begin{eqnarray*}
({\boldsymbol{x}}'_1, {\boldsymbol{x}}'_2) = 
({\boldsymbol{x}}_1, {\boldsymbol{x}}_2) G^t ,  ~~~
({{\varphi}}'_1, {{\varphi}}'_2) = ({{\varphi}}_1, {{\varphi}}_2) G^{-1} ,\\
G=\left(
\begin{array}{cc}
p & 0 \\
q & kq 
\end{array}
\right),  ~~~~~~
U ~\to ~ U'= \frac{1}{2}\left(
\begin{array}{cc}
\dfrac{1}{|{\boldsymbol{x}}'_1|}&0\\
0&\dfrac{1}{|{\boldsymbol{x}}'_2|}
\end{array}
\right). 
\end{eqnarray*}
The effect of adding $N_F$ flavors is to include 
$\Delta U
=\frac{1}{2}\left(
\begin{array}{cc}
0&0\\
0&\dfrac{N_F }{|{\boldsymbol{x}}_2 |}
\end{array}
\right)$ to $U$ and thus 
\begin{eqnarray*}
\Delta U'
=\frac{1}{2}\left(
\begin{array}{cc}
\dfrac{q N_F }{kp\SL}
&\dfrac{- N_F }{k\SL}\\
\dfrac{- N_F }{k\SL}
&\dfrac{p N_F }{kq\SL}
\end{array}
\right),  ~~~
\SL=|p{\boldsymbol{x}}'_2 - q{\boldsymbol{x}}'_1|.
\end{eqnarray*}
We see that due to non-zero $N_F$, 
$({\boldsymbol{x}}'_1, {\boldsymbol{x}}'_2)$ 
should be rotated simultaneously by a common element of 
$SO(3)_R$ in order to preserve $\SL$. Moreover, $U(1)_b \times U(1)_d$ corresponds to two $U(1)$'s of 
$(\varphi'_1 ,\varphi'_2 )$ which 
can be promoted to a 
local symmetry and 
offset by gauge transformations of $(A'_1, A'_2)$. 
These together again agree with the above argument.

In order to make the structure of Eschenburg space, 
through 
${\boldsymbol{x}}'_1\to -{\boldsymbol{x}}'_1$, 
${{\varphi}}'_1 \to -{{\varphi}}'_1$ and 
rewriting $\Delta U'$ as 
\begin{eqnarray*}
\Delta U'= \frac{1}{2} \left(
\begin{array}{cc}
\dfrac{t_1^2 }{t_3 |t_1 {\boldsymbol{x}}'_1 + t_2 {\boldsymbol{x}}'_2| }
&\dfrac{t_1 t_2  }{t_3 |t_1 {\boldsymbol{x}}'_1 + t_2 {\boldsymbol{x}}'_2| }\\
\dfrac{t_1 t_2  }{t_3 |t_1 {\boldsymbol{x}}'_1 + t_2 {\boldsymbol{x}}'_2| }
&\dfrac{t_2^2}{t_3 |t_1 {\boldsymbol{x}}'_1 + t_2 {\boldsymbol{x}}'_2| }
\end{array}
\right), 
\end{eqnarray*}
we then find that 
$ds^2_{8D}$ in \eqref{8} 
leads to Eschenburg space labeled by 
three coprime 
natural numbers $(t_1, t_2, t_3)=(qN_F,pN_F,kpq)$.

When $t_1 \ne t_2 \ne t_3$, it has the least isometry 
$SU(2)\times U(1)^2$ and 
preserves 
a fraction $3/16$ of 32 SUSY (defining 
feature of a cone over 7D tri-Sasakian manifolds)%
\footnote{An enhancement 
to a fraction 3/8 happens while one zooms into 
the 
near-horizon region of M2-branes, i.e. $R^{1,2}\times 
$ Cone$({{\mathcal B}_7}) \to AdS_4 \times {{\mathcal B}_7}$.}.
According to \cite{Lee:2006ys}, 
one has the following relation between 
5- and 7-cycles inside Eschenburg space: 
\begin{eqnarray}
\frac{{{vol}}({S^5})}{vol({{\Sigma}_5})}=\frac{{{vol}}({S^7})}{vol({\tilde{{\mathcal M}}_7})}
=\frac{(q+p )(N_F+kq)(N_F+kp)}{\big( N_F + k(q  + p)\big)}.
\label{vol}
\end{eqnarray}
\subsection{Entropy from M-theory dual}
Having said that the transverse geometry is a 
8D hyperK{\"a}hler cone, after the 
backreaction of M2-branes we are left with 
$AdS_4 \times \tilde{\mathcal M}_7$  
under 
the normalization 
\begin{eqnarray}
&6R^6 vol(\tilde{\mathcal M}_7)=(2\pi \ell_p)^6 N, \quad\quad
&R_{S^7}^6=2^5\pi^2 Nl_p^6.
\label{NOR}
\end{eqnarray}
Note that $R=2R_{AdS}$ is the radius of 
$\tilde{\mathcal M}_7$. This background will be taken 
as the cornerstone of studying the strongly coupled behavior of our ${\mathcal N}$=3 CSM theory.  
To count degrees of freedom via the above M-theory dual, 
we replace $AdS_4$ with $AdS$-Schwarzschild black hole and 
evaluate its Bekenstein-Hawking entropy.

$AdS$-Schwarzschild black hole metric is given by 
\begin{eqnarray}
ds^2=\left(\dfrac{4r^2}{R^2}+1-\dfrac{M}{r}\right)d\tau^2 +\dfrac{dr^2}{\left(\dfrac{4r^2}{R^2}+1-\dfrac{M}{r}\right)}+r^2d\Omega^2_{2}.
\label{ADB119}
\end{eqnarray}
This metric can serve as a dual description of the 
finite-temperature CSM theory on $S^1\times S^2$. 
\eqref{ADB119} is smooth if the period of $\tau$ 
satisfies 
\begin{eqnarray} 
&\beta = \dfrac{\pi R^2r_0}{3r_0^2+\frac{R^2}{4}}
\label{BE13}
\end{eqnarray}
where $r_0$ is the horizon radius. 
Solving \eqref{BE13} in terms of $\beta$, we obtain 
\begin{eqnarray}
&r_0=\dfrac{\pi R^2}{6\beta}+\sqrt{\left(\dfrac{\pi R^2}{6\beta}\right)^2-\dfrac{R^2}{12}}. \label{HO121}
\end{eqnarray}
From \eqref{HO121}, it is found that 
$AdS$ black holes exist when $\beta < \pi {R}/{\sqrt{3}}$ 
and Hawking-Page phase transition \cite{Wi1} occurs at $\beta_c=\pi {R}/{2}$ above the temperature bound.

We can use Bekenstein-Hawking area law to yield 
the entropy per $\frac{1}{4}vol(S^2) R^2$: 
\begin{eqnarray}
&S\equiv\dfrac{2^{\frac{3}{2}}\pi^2N^{\frac{3}{2}}}{27\beta^2}\left(1+\sqrt{1-\dfrac{3\beta^2}{\pi^2R^2}}\right)^2\sqrt{\dfrac{vol(S^7)}{vol({\tilde{{\mathcal M}}_7})}} \label{ENT16} \nonumber\\
&\to \dfrac{2^{\frac{7}{2}}\pi^2N^{\frac{3}{2}}}{27\beta^2}
\sqrt{\dfrac{vol(S^7)}{vol({\tilde{{\mathcal M}}_7})}}\quad (\text{at high temperature}). \label{ENT17}
\end{eqnarray}
Note that 
the unit volume of 7-cycle of Eschenburg space is related 
to that of $S^7$ via \eqref{vol}. See Appendix A for 
another point of view on the derivation of $S$ from 
GKP-W relation. 
By assuming $N_F\ll k$ $(\lambda \ll N/N_F)$ and expanding 
\eqref{ENT16} in powers of $1/k$, \eqref{ENT17} looks like 
 \begin{eqnarray}
&S=\dfrac{2^{\frac{7}{2}}\pi^2 N^{\frac{3}{2}}}{27\beta^2}
\Biggl[\sqrt{kpq}
+\dfrac{N_F\left({p}^{2}+{q}^{2}+pq \right)}{2\sqrt{kpq}(p+q)}+ \nonumber \\
&+ \dfrac{N_F^2}{(kpq)^{\frac{3}{2}}}\left(-\dfrac{1}{2}\left({p}^{2}+{q}^{2}\right)+\dfrac{3\left( {p}^{2}+{q}^{2}+pq \right)^2}{ 8\left( p+q \right) ^{2}} \right) + {\mathcal O}(N_F^3k^{-\frac{5}{2}})\Biggr]. \label{FEN19}
\end{eqnarray}
It is convenient to rewrite \eqref{FEN19} in terms of 
't Hooft coupling as 
\begin{eqnarray}
&S=\dfrac{2^{\frac{7}{2}}\pi^2 }{27\beta^2}
\Biggl[N^2 \sqrt{\dfrac{pq}{\lambda}}  
+\dfrac{\sqrt{\lambda}N N_F\left({p}^{2}+{q}^{2}+pq \right)}{2\sqrt{pq}(p+q)} \nonumber \\
&+ \dfrac{\lambda^{\frac{3}{2}}N_F^2}{(pq)^{\frac{3}{2}}}\left(-\dfrac{1}{2}\left({p}^{2}+{q}^{2}\right)+\dfrac{3\left( {p}^{2}+{q}^{2}+pq \right)^2}{ 8\left( p+q \right) ^{2}} \right)\Biggr]+.... 
\label{FEN20}
\end{eqnarray}
Setting $p=q=1$ in \eqref{FEN20}, we recover 
results of the flavored 
ABJM theory in \cite{Gaiotto:2009tk}. 
The 1st term on RHS of \eqref{FEN19} is the famous 
$N^{\frac{3}{2}}$ factor of M2-branes. The 2nd term 
can be interpreted as the tree-level effect of adding flavors 
as will be shown to be captured by IIA probe D6-branes. 
The 3rd term 
proportional to $\lambda^{\frac{3}{2}}N_F^2$ represents 
degrees of freedom from mesonic flavor states. 
Higher order terms may describe the interaction 
between flavor and bi-fundamental fields. We also find the 2nd term in \eqref{FEN20} $\propto \sqrt{\lambda} N N_F$ 
has an additional $\sqrt{\lambda} \gg 1$ factor compared to weak coupling results. This may suggest that 
in strong coupling regime degrees of freedom due to flavors gets 
increasing quite a lot.

\subsection{On-shell action of flavor D6-brane }
We now turn to Type IIA viewpoint of evaluating the entropy. 
This involves treating flavors as probe branes in a given geometry. 
To discuss their on-shell action, one must first clarify 
how they are embedd. 

Let us briefly describe 
the dual geometry of ${\mathcal N}$=4 SCFT 
\cite{Imamura:2008ji} constructed via 
IIB $N$ circular D3-, $p$ NS5- and 
$q$ $(1,k)$5-branes: 
\begin{eqnarray}
ds^2_{11D} = \frac{R^2}{4} ds^2_{AdS_4} + R^2 ds_7^2 , 
~~~ R=\ell_p (2^5 N kpq \pi^2)^{1/6},\nonumber\\
ds_7^2 = 
d\xi^2 + \frac{1}{4}\cos^2\xi \Big( (d\chi_1+ 
\cos\theta_1 d\phi_1)^2 + 
d\theta_1^2 + \sin^2\theta_1 d\phi_1^2 \Big) \nonumber\\
+ \frac{1}{4}\sin^2\xi \Big( (d\chi_2+ 
\cos\theta_2 d\phi_2)^2 + d\theta_2^2 + \sin^2\theta_2 
d\phi_2^2 \Big),\nonumber\\ 
(\chi_1, \chi_2) \sim (\chi_1+ \frac{4\pi}{kp},
\chi_2 + \frac{4\pi}{kq} ) \sim 
(\chi_1+ \frac{4\pi}{p}, \chi_2), \nonumber\\ 
0< \xi \le \frac{\pi}{2},~~~ 0< \theta_i \le \pi, ~~~
0<\phi_i \le 2\pi.  
\label{Y7}
\end{eqnarray}
Its isometry, two copies of 
$SU(2)\times U(1)$, is easily read off 
because there are 
two (orbifolded) $S^3$ fibered over a segment $[0,1]$. 
It is straightforward to show that \eqref{Y7} is equivalent to 
\eqref{8} with \eqref{u} via including the near-horizon warp factor of M2-branes and changing variables as in \cite{Hikida:2009tp}.

For simplicity, we set $p=q$ and 
KK reduce to IIA string theory. Recall 
\begin{eqnarray*}
ds_{11D}^2 = e^{-\frac{2}{3}\Phi} ds^2_{IIA} + 
 e^{\frac{4}{3}\Phi} (d\varphi+\cdots)^2, 
\end{eqnarray*}
then, 
\begin{eqnarray}
&ds_7^2 = ds^2_6 + \dfrac{1}{k^2 q^2}(d\tilde{y}+
\tilde{A})^2,~~~~~e^{2\Phi}=\dfrac{R^3}{k^3 q^3}, \nonumber\\
&\tilde{A}=
kq\Big( \dfrac{1}{2}\cos^2\xi ( d\psi + 
\cos\theta_1 d\phi_1 ) + \dfrac{1}{2}\sin^2\xi \cos\theta_2 
d\phi_2 \Big), \nonumber\\
&ds^2_6 =
d\xi^2 + \dfrac{1}{4}\cos^2\xi \sin^2\xi \Big( d\psi + 
\cos\theta_1d\phi_1 -\cos\theta_2 d\phi_2 
\Big)^2 \nonumber\\
&+ \dfrac{1}{4}\cos^2\xi (d\theta_1^2 + \sin^2\theta_1 d\phi_1^2)+ \dfrac{1}{4}\sin^2\xi (d\theta_2^2 + \sin^2\theta_2 
d\phi_2^2), \nonumber\\
&ds^2_{IIA}=L^2 (ds^2_{AdS4_4} + {4}ds^2_{6}),~~~~~~
L^2 = \dfrac{R^3}{4kq},\nonumber\\
&0< y=\dfrac{1}{kq}\tilde{y}\le \dfrac{2\pi}{kq},~~~ 0< \psi \le \dfrac{4\pi}{q},~~~
\chi_1 =  \psi + 2 y ,~~~\chi_2=2y
\label{IIA} 
\end{eqnarray}
with fluxes 
\begin{eqnarray}\label{F2kq}
 F_2=kq\Big(
\cos\xi \sin\xi d\xi \wedge (d\psi+ 
\cos\theta_1d\phi_1-\cos\theta_2 d\phi_2)\nonumber\\
 -\frac{1}{2} \cos^2\xi 
\sin\theta_1 d\phi_1\wedge d\theta_1 
-\frac{1}{2} \sin^2\xi 
\sin\theta_2 d\phi_2\wedge d\theta_2 \Big)=-\frac{kq}{2L^2}\omega_2,\nonumber\\
F_4=-\frac{3}{8}R^3 \epsilon_{AdS_4}, ~~~
~~~
\epsilon_{AdS_4}=r^2 dt\wedge
 dx^{1} \wedge dx^{2} \wedge dr. 
\end{eqnarray}
It is obvious that 
$\mathcal{N}$=6 ABJM case differs from 
ours only by a factor $q$. As will be explained more 
in Appendix A, 
by imposing ${ \boldsymbol x}_2=(789)=0$ (locus of IIB flavor 
D5-branes) on \eqref{Y7}, there then appears 
$AdS_4 \times S^3/{\bf Z}_{2q}$: 
\begin{eqnarray}
&\psi '=\psi/2, ~~~\theta_1=\theta_2 (=\theta),~~~ \phi_1=-\phi_2 (=\phi),~~~ \xi=\dfrac{\pi}{4},  \nonumber \\
&ds^2_{S^3/Z_{2q}}=\dfrac{1}{4}\left(d\psi '+\cos\theta d\phi\right)^2+\dfrac{1}{4}(d\theta^2+\sin^2\theta d\phi^2). 
\label{SPE32}
\end{eqnarray}

Let us go to evaluate the on-shell action 
of probe D6-branes. 
The induced metric at finite temperature is   
\begin{eqnarray}\label{ME33}
&ds^2_{D6}=L^2\left(\dfrac{dr^2}
{r^2 \big(1-( \dfrac{r_0}{r})^3 \big)}-r^2 \big(1-(\dfrac{r_0}{r})^3 \big) dt^2+ r^2 d\vec x^2  +  4ds^2_{S^3/Z_{2p}}\right),
\end{eqnarray}
where Hawking temperature is given by $T={3r_0}/{4\pi}$. 
The on-shell action of $N_F$ D6-branes 
per volume $V_2$ of $\vec{x}=(x^1, x^2)$ is 
\begin{eqnarray}
I_{D6}=-\dfrac{2^3 N_F}{(2\pi )^6}e^{-\Phi}L^7 vol(S^3/{\bf Z}_{2p})\cdot \int dt \int ^{\infty}_{r_0}dr ~r^2 ~\to ~\dfrac{2^{\frac{7}{2}}}{81}\sqrt{\lambda}\pi^2 N_F NT^2,
\label{SD6}
\end{eqnarray}
where we have subtracted the divergent part at infinity. 
Note that $I_{D6}$ does not 
depend on $p$ (or $q$). From \eqref{SD6} the free energy 
and entropy per $V_2$ are 
\begin{eqnarray}
F_{D6}=-TI_{D6}=-\dfrac{2^{\frac{7}{2}}}{81}\sqrt{\lambda}\pi^2 N_FNT^3,\quad ~~~~
S_{D6}=\dfrac{2^{\frac{7}{2}}}{27}\sqrt{\lambda}\pi^2 N_FNT^2.
\label{EF23}
\end{eqnarray}
Interestingly, we find that $S_{D6}$ is again 
accompanied by a $\sqrt{\lambda}$ 
factor compared to zero-coupling approximation. In addition, $S_{D6}$ is 
larger than  
the 2nd term in 
\eqref{FEN19} by a factor $4/3$ if $p=q$! 
But we should be cautious because 
$\lambda$ has different values in the two cases. 
In M-theory where M-circle ($\sim R/\ell_p kq \gg 1$) is decompactified, $\lambda \gg k^4 q^4$. 
Here, in IIA theory $g_s \ll 1$ means that 
$1 \ll \lambda \ll k^4 q^4$. 
It still seems interesting to pursue this $4/3$ problem against 
the famous one in \cite{Gubser:1996de}.

\section{Superconformal index}

Superconformal indices of 3D SCFTs are 
considered in \cite{Bhattacharya:2008bja, MIN, 
Dolan:2008vc, Choi:2008za, Kim:2009wb}. 
Let us compute that of our 
${\mathcal N}$=3 
SCFT containing flavors. 
Because the internal 
7-manifold $\tilde{{\mathcal M}}_7$ is not homogeneous in general, 
to study KK spectra on $\tilde{{\mathcal M}}_7$ is quite 
difficult. 
As prescribed in \cite{Bhattacharya:2008bja},  
\begin{eqnarray*}
I = \Tr (-)^F x^{\epsilon+ j }y_1^{h_2} \cdots  y_{M-1}^{h_M}
\end{eqnarray*} 
receives contributions from short multiplets. 
$M=[{\mathcal N}/2]$ gets related to its 
superconformal group $OSp({\mathcal N}|4)$. 
Also, $\epsilon$, $j$ and $h_i$ are 
eigenvalues of Cartan generators of bosonic subgroup 
$SO(2)\times SO(3)\times SO({\mathcal N})$ of $OSp({\mathcal N}|4)$. 
For ${\mathcal N}$=3, $I$ gets simplified to 
\begin{eqnarray}
I = \Tr (-)^F x^{\epsilon+ j }. 
\label{de}
\end{eqnarray} 
Using an unitary matrix model prescribed in \cite{A1,Choi:2008za}, 
we can instead compute \eqref{de} by 
\begin{eqnarray*}
I=\int \prod^{2q}_{I=1} DU_I \exp \Big( 
\sum_R \sum^{\infty}_{n=1} \dfrac{1}{n} F_R (x^n) \chi_R (U_I^n)
\Big)\\
=\int \prod^{2q}_{I=1} DU_I 
\exp  \sum^q_{i=1} \sum^{\infty}_{n=1} 
\dfrac{1}{n} \Big( F^t_n \big(  \Tr (U^n_{2i})  \Tr (U^{-n}_{2i+1}) + (n \leftrightarrow -n)  \big) \\
+ F^{unt}_n \big( \Tr (U^n_{2i-1})  \Tr (U^{-n}_{2i}) + (n \leftrightarrow -n) \big)  \\
+ N_F^{2i} F^{f}_n \big(  \Tr (U^n_{2i}) + \Tr (U^{-n}_{2i}) \big) 
+ N_F^{2i-1} F^{f}_n  \big( \Tr (U^n_{2i-1}) + \Tr (U^{-n}_{2i-1}) \big) \Big) 
\end{eqnarray*} 
where all 
conventions about unitary matrices follow \eqref{ZZ}. 
Again, we assumed that NS5- and $(1,k)$5-branes are
placed pairwise on the circle such that 
$p=q$. 
By using same techniques as in \eqref{ZZ}, 
\begin{eqnarray*}
I=\int \prod_{i,n} d\rho_{i,n} d\chi_{i,n} \exp \sum_{i,n} -\dfrac{N^2}{n}
\Big(  |\rho_{i,n}|^2 +  |\chi_{i,n}|^2    \\
- \dfrac{1}{N} N_F^{2i} F^{f}_n ( \rho_{i,n} +  \rho_{i,-n} ) 
- \dfrac{1}{N} N_F^{2i-1} F^{f}_n  ( \chi_{i,n} + \chi_{i,-n} ) 
\\
- F_n^{unt} \big(  \chi_{i,n} \rho_{i+1,-n} + (n \leftrightarrow -n) \big) 
-{F}_n^t \big( \rho_{i,n} \chi _{i,-n} + (n \leftrightarrow -n) \big)  \Big)
\end{eqnarray*}
with 
\begin{eqnarray*}
F_n^{unt} ={F}_n^t =F^f_n=F_n =F(x^n), \quad \quad
F(x)=\dfrac{\sqrt{x}}{1+x}.
\end{eqnarray*}

Further setting $N_F^i=m$ for simplicity and rewriting $I$ as 
($M_n$: $4q \times 4q$ matrix)
\begin{eqnarray*}
I=\int \prod_{n} dc_n \exp \sum_{n} -\frac{N^2}{2n} 
( c_n^t M_n c_n + p_n^t c_n + c_n^t p_n ),\nonumber \\
c_n^t= (\rho_{1,n}, 
\chi_{1,n}, \rho_{1,-n}, \chi_{1,-n},\cdots), \quad \quad
p_n=-\frac{2m}{N}F^f_n {\bf 1}_{1\times q}, 
\end{eqnarray*}
one soon performs 
this Gaussian integral to yield 
\begin{eqnarray}
I=\prod^\infty_{n=1}  
\dfrac{(1+x^{n})^{2q}}{(1-x^{nq})^2} \cdot 
\exp \sum_{n} {\mathcal K}_n,\quad \quad
{\mathcal K}_n = \dfrac{N^2}{2n} p_n^t M_n^{-1 } p_n = 
\frac{2m^2}{n} F_n^2
\sum_{a,b} \big( M_n^{-1 }\big)_{a b}
\end{eqnarray}
with 
\begin{eqnarray}
M_n= \begin{pmatrix}
  \ddots &  & {\mathcal Q}^T_{4\times 4} \\
 & {\mathcal S}_{8\times 8} &   \\
 {\mathcal Q}_{4\times 4} &  & \ddots
 \end{pmatrix}
, \quad 
 {\mathcal S}_{8\times 8}=
 \begin{pmatrix}
  {\mathcal R}_{4\times 4}  & {\mathcal Q}_{4\times 4} \\
     {\mathcal Q}^T_{4\times 4}     &    {\mathcal R}_{4\times 4}
 \end{pmatrix}.
\end{eqnarray}
Here, $\ddots$ denotes ${\mathcal S}$ and $\mathcal{Q}$ in the corner of $M_n$ is the contribution from the periodic condition $\rho_{q+1,n}=\rho_{1,n}$. ${\mathcal Q}$ has only non-zero elements $\mathcal{Q}_{41}=\mathcal{Q}_{23}=-F_{n}$, while
\begin{eqnarray*}
{\mathcal R}= 
 \begin{pmatrix}
0& {\mathcal P} \\
{\mathcal P}&0
 \end{pmatrix}, \quad \quad
 {\mathcal P}= 
  \begin{pmatrix}
1& -F_n \\
-F_n &1
 \end{pmatrix}.
 \end{eqnarray*} 
To evaluate how the $flavor$ sector contributes to 
$I$ lies in expanding the exponential w.r.t. $m$. 
We leave the comparison with gravity results
in future works. 
\section{Baryonic operator}
Let us examine the correspondence concerning 
baryonic operators. 
If we assume $N \gg N_F$, baryons 
like $\epsilon^{i_1 \cdots i_N} Q^I \cdots Q^I$ must 
be ruled out. Then, the possibility lies in 
\begin{eqnarray}
B^I=\epsilon_{j_1 \cdots j_N} 
\epsilon^{i_1 \cdots i_N} A^I \cdots A^I , \quad \quad 
B^J=\epsilon_{j_1 \cdots j_N} 
\epsilon^{i_1 \cdots i_N} A^J \cdots A^J 
\label{BB}
\end{eqnarray} 
where $SU(2)_R$, color and flavor 
indices are suppressed, while 
$I$ ($J$) stands for twisted (untwisted) 
hypermultiplets. Their conformal dimensions can be 
determined from the superpotential 
${\mathcal L}'_{pot}$ to be $\Delta(B)={N/2}$. 
On the gravity side, $\Delta$ 
can be confirmed via 
M5-branes wrapping 
5-cycles $\Sigma_5$ inside Eschenburg space. Upon 
using \eqref{vol} and \eqref{NOR}, 
\begin{eqnarray*}
\Delta=R_{AdS} \cdot m_{M5}=\frac{1}{2} \tau_{M5}R^6 vol (\Sigma_5)=
\frac{\pi N}{6} \frac{vol(\Sigma_5)}{vol(\tilde{{\mathcal M}}_7)} =\frac{N}{2}\\
\end{eqnarray*}
for large M5-brane mass. We see that $\Delta$ is independent of 
($t_1, t_2, t_3$) as pointed out in \cite{Lee:2006ys}. 

When it comes to degeneracy, both di-baryons above 
having $N+1$ degeneracy form 
a spin ${N}/{2}$ rep. of $SU(2)_R$. 
From 
\begin{eqnarray*}
\frac{vol(\tilde{{\mathcal M}}_7)}{vol(\Sigma_5)}
 =\frac{vol(S^7)}{vol(S^5)}=\frac{\pi}{3} \sim 
\frac{1}{12} vol (S^2),
\end{eqnarray*}
we can think that topologically what is transverse to an 
M5-brane inside $\tilde{{\mathcal M}}_7$ 
is roughly a 2-sphere such that the argument similar 
to \cite{GK} is still applicable.  
That is, the degeneracy of di-baryons is accounted for by 
$N$ units of 7-form flux penetrating $S^2$. 
Collective coordinates of an M5-brane thus behave quantum  mechanically as if 
there were $N+1$ degenerated states in the 
lowest Landau level under $N$ units of 
magnetic flux through $S^2$.

Finally, we comment on how many 
independent di-baryons are there. 
According to \cite{Imamura:2008ji} without flavors, 
the decomposability of 
$\prod_I B^I$ and 
$\prod_J B^J$ (dressed by appropriate monopole operators) into 
mesons gives rise to totally $p+q-2$ independent di-baryons. 
A detailed survey on the homology $H_5 ({\mathcal M}_7,{\bf Z})={\bf Z}^{p+q-2}$ reveals the same thing%
\footnote{See \cite{Imamura:2009ph} for detailed considerations about homology in 
AdS$_4$/CFT$_3$.}. These can be put another way, i.e. 
gauge-variant di-baryons are 
charged under $p+q-2$ $U(1)$ gauge fields except for 
the two (diagonal one and dual photon) which are not involved in performing a hyperK{\"a}hler 
quotient as said in section 1. 
Naively, this means that the RR 6-form potential 
should therefore be expanded 
like $C_6 \sim \sum^{p+q-2}_{I=1} \omega^I_5 \wedge A^I$. 
In other words, 
$H_5 ({{\mathcal M}}_7,{\bf Z})
={\bf Z}^{p+q-2}$ ($\omega_5$: volume form of 5-cycle).

In our case, given 
Betti numbers 
\begin{eqnarray*}
b_2(\tilde{{\mathcal M}}_7)=b_5 (\tilde{{\mathcal M}}_7)=1, 
\end{eqnarray*}
it seems there is only 
one single di-baryon though. 
In view of \eqref{BB}, it seems there should be more 
independent di-baryons according to arguments given above. 
We wish to resolve the discrepancy in a future work.

\section{Conclusion}

In this note we provide a gravity dual for the flavored ${\mathcal N}$=4 Chern-Simons-matter theory which is a kind of 
${\mathcal N}$=3 SCFT. 
From the following three viewpoints: \\
1. SUSY and global symmetry match\\
2. hyperK{\"a}hler quotient construction of the moduli space\\ 
3. GGPT method of identifying the M-theory transverse geometry 
from the given Type IIB 5-brane configuration\\
we get confident in regarding our proposed 
Eschenburg space as an adequate candidate.

To study further the 
correspondence between both, 
we go to count degrees of freedom. On the field theory side, 
this is done by taking large $N$ zero-coupling approximation such that 
an unitary matrix model previously known fulfills our purpose. 
On the gravity side, two approaches are tried, 
namely, we calculate the entropy from both the 
11D $AdS$-Schwarzschild-Eschenburg black hole 
and an on-shell action of probe D6-branes in 
Type IIA geometry which is dual to ${\mathcal N}$=4 CSM. 
It is seen that 
field theory results are corrected 
by multiplying a factor 
$\sqrt{\lambda}$ to $NN_F$ terms. 
This suggests that in strong coupling regime degrees of freedom 
due to adding 
flavors increase extremely. 
We also study gravity duals of 
mesonic and baryonic operators and 
find agreements on their 
conformal dimensions, and so on.

Moreover, an ${\mathcal N}$=3 superconformal index is computed, though a comparison with 
the one from gravity is left in 
a future work due to essential 
difficulties in deriving Kaluza-Klein 
spectra on inhomogeneous $\tilde{{\mathcal M}}_7$. 
Nevertheless, for $(t_1, t_2, t_3)=(1,1,1)$, i.e. $\tilde{{\mathcal M}}_7=N(1,1)$, its KK spectra are known in some literature
\cite{Termonia:1999cs, Fre':1999xp, Billo:2000zr} and 
hence the comparison seems worthy of trying. 
Since Eschenburg space metric has not yet been fully 
exploited, we wish to report progress towards its 
application soon.

\section*{Acknowledgements}
T.S.T. is grateful to 
Satoshi Yamaguchi for helpful comments and Yosuke Imamura for 
a series of valuable discussions on 
${\mathcal{N}}$=4 CSM theory. M.F. thanks Yasuaki Hikida and 
Hiroshi Ohki for helpful advice. T.S.T is supported in part by  the postdoctoral program at RIKEN.

\appendix 
\section{Degrees of freedom in 3D $\mathcal{N}=3$ SCFT}
We can roughly evaluate the degrees of freedom of the strongly coupled 3D SCFT 
via GKP-W relation 
\cite{Gubser:1998bc, Witten:1998qj}%
\footnote{This part is inspired by the 
lecture note of Yosuke Imamura.}. 

We just compute the correlation function of two energy-momentum tensors 
\begin{eqnarray*}\label{langT}
\langle T(x) T(y)\rangle=\frac{\delta^2 S_{gravity}}{\delta h \delta h}\sim \frac{c}{|x-y|^{2\Delta}}, \quad \quad \Delta=3
\end{eqnarray*}
where $h$ is the metric 
perturbation around $AdS_4$ boundary, while $c$ may contain the information about degrees of freedom in the 3D SCFT. 
Because $c$ is dimensionless and ($G_D$: Newton constant)
\begin{eqnarray*}\label{sgraf}
 S_{gravity} =\frac{1}{G_4} \int d^4 x \sqrt{-g} 
({\mathcal R} - \Lambda), 
\end{eqnarray*}
the only choice for $c$ is  
\begin{eqnarray*}\label{csim}
c\sim \frac{R^2_{AdS}}{G_{4}}, \quad\quad G_4=\frac{G_{11}}{R^7 vol (\tilde{{\mathcal M}}_7)} ,\\
R=2R_{AdS},\quad\quad G_{11}=(2\pi)^8 \ell_p^9, 
\quad\quad 
\to c\sim \frac{N^{\frac{3}{2}}}{\sqrt{vol 
(\tilde{{\mathcal M}}_7)}}
\end{eqnarray*}
where $R$ is the radius of $\tilde{{\mathcal M}}_7$ and we have used \eqref{NOR}. 
As is shown in section 4, $c$ is exactly 
what is computed via Bekenstein-Hawking area law via 
its M-theory dual.

\section{Taub-NUT space}
In this Appendix, we show that probe D6-branes 
wrap 012 plus 
4D Taub-NUT space inside the 7D cone in IIA theory.

Recall that 
flavor D5-branes 
occupying 
$(012345)$ are localized at 
${\boldsymbol x}_2=(789)=0$. 
Regarding this constraint as a moment map (at zero level set), we can further 
perform a hyperK{\"a}hler quotient via rearranging 
$ds^2_{8D}$ into $(\varphi_2$: M-circle, $U=\frac{1}{2}\tilde{U}$)
\begin{eqnarray}
ds^2_{8D}
=\frac{1}{4}\Big( \tilde{U}_{ij} 
d{ \boldsymbol{x}}_i \cdot  d {\boldsymbol{x}}_j + 
 \frac{4}{ \tilde{U}_{11}} (d\varphi_1 + A_1)^2  \Big) 
 + \frac{\tilde{U}_{11}}{\det \tilde{U}} \Big(d\varphi _2+A_2-\frac{\tilde{U}_{12}}{\tilde{U}_{11}}(d\varphi_1 +A_1)\Big)^2.
\end{eqnarray}

A rescale is done to get a period $2\pi$ M-circle.  
Imposing 
${\boldsymbol x}_2=0$ and throwing away the last term, 
we have ($p=q$, ${ \boldsymbol x}_1={ \boldsymbol \rho},~ \rho=|{ \boldsymbol \rho}|, ~
\varphi_1=\psi \in (0,4\pi]$)  
\begin{eqnarray}
&ds_{4}^2 = \dfrac{1}{2}\Big(
\dfrac{q}{\rho}  d{ \boldsymbol \rho}^2 + 
\dfrac{\rho}{q} (d\psi + A_1 )^2  \Big),\nonumber\\
&A_1 \to 2q\dfrac{-\rho^1 d \rho^2 + \rho^2 d\rho^1} {\rho(\rho 
+ \rho^3) }
= -2q{ \boldsymbol \omega}\cdot d { \boldsymbol \rho}, ~~~~
\nabla \times { \boldsymbol \omega} = -\nabla\dfrac{1}{\rho},
\label{crho}
\end{eqnarray}
which represents a multi-centered Taub-NUT 
whose $2q$ NUTs coincide. Owing to 
the cone structure, 
making M2-branes backreact and taking near-horizon limit, 
we have constant dilaton field 
\begin{eqnarray*}
e^{\frac{4}{3}\Phi}=H^{\frac{1}{3}}
(r^2=|{\boldsymbol x}'_1|+|{\boldsymbol x}'_2|)\cdot
\big( 
\dfrac{\tilde{U}_{11}}{\det \tilde{U}}\big)=
{\text {const.}},\\
H=1 + \dfrac{\ell^6_p 2^5 N'\pi^2  }{ r^6} 
\sim (\dfrac{R^2}{r^2})^3 \quad\quad r\to 0,
\end{eqnarray*}
which promises an $AdS$ factor. 
Therefore, 
one can finally arrive at flavor D6-branes with  
worldvolume $AdS_4\times S^3/{\bf Z}_{2q}$:   
\begin{eqnarray}\label{TN}
ds^2_{D6}=L^2 \big( ds_{AdS_4}^2+4ds^2_{S^3/Z_{2q}} \big).
\end{eqnarray}

On the other hand, 
if the level set is non-zero ${ \boldsymbol x}_2 =\xi\ne 0 $, 
i.e. adding massive fundamental flavors, 
it is readily seen that all NUTs will not coincide and 
$TN_{4}$ is partially resolved. 

\section{Meson spectrum}
Here, we consider mason spectra from Type IIA geometry in \eqref{IIA}. 
This involves a D6-brane embedding with worldvolume action described by ($\ell_s =1$)
\begin{eqnarray}\label{D6action}
 S_{D6} = - T_{D6} \int d^{7}x ~ \sqrt{ - \det ( g_{ab} + B_{ab}+2 \pi F_{ab}) }
    -T_{D6} \int e^{2\pi F+B} \wedge \sum_{p} C_p.
\end{eqnarray}
We take the static gauge such that its worldvolume is 
parameterized by 
$(t,x,y,r,\theta,\phi,\psi')$ with 
$\theta=\theta_1=\theta_2$ and $\phi=\phi_1=-\phi_2$. 
The scalar perturbation on a stack of D6-branes concerning meson spectra is $\delta \xi=\eta = \rho(r) e^{ip\cdot x} Y_\ell (\Omega)$. 
Its angular part can be expanded by spherical harmonics 
on $S^3$:  
\begin{eqnarray}\label{nabY}
\nabla^2 Y_\ell (\Omega) = -\ell (\ell+2) Y_\ell (\Omega).
\end{eqnarray}
Let us assume in Type IIB picture there are totally $F$ flavor D5-branes distributed over $2q$ intervals of $x^6$ as 
$\Big( F_1,\cdots,F_{2q} \Big)$ $(F=\sum^{2q}_{I=1} F_I)$ such that the flavor symmetry gets broken 
like $U(F)\to \prod_I U(F_I)$. Note that via T-dualizing 
$x^6$ to IIA the above information is encoded in 
the following holonomy (Wilson loop) 
\begin{eqnarray}\label{expi}
\exp ({i\oint A_{\psi'}} d\psi') =\bigoplus^{2q}_{I=1} \omega^I {\bf 1}_{F_I }
\end{eqnarray}
on $F$ D6-branes 
with $\omega=\exp ({2\pi i /2q})$ due to $\pi_1 (S^3/{\bf Z}_{2q})={\bf Z}_{2q}$ in \eqref{TN}.

Moreover, spherical harmonics can be labeled 
by $Y^{IJ}_\ell (\Omega)$, i.e. 
the open string scalar mode $\eta$ can be either of 
adjoint rep. ($I=J$) or 
bi-fundamental rep. ($I\ne J$) w.r.t. flavor 
groups depending on on which two 
stacks of 
D6-branes its ends are. Because 
only modes 
surviving the projection $\Gamma =\omega^{I-J} \exp 
\dfrac{4\pi i J^3_L}{2q}$ ($J_L$ is the generator of $SO(4)\simeq SU(2)_L \times SU(2)_R$ of $S^3$ and 
$\omega^{I-J}$ 
stands for the acquired holonomy) 
remain, therefore 
\begin{eqnarray}\label{2mL}
2 m_L + I-J \in 2q{\bf Z}. 
\end{eqnarray}

For $Y_{\ell}$ of 
$(m_L, m_R) =({\ell}/{2}, {\ell}/{2})$, this implies 
\begin{eqnarray}\label{ellk}
\ell =k +  2q{\bf Z},~~~~~~k=0, \cdots, 2q-1. 
\end{eqnarray}
Furthermore, due to \cite{Hikida:2009tp} 
\begin{eqnarray*} 
\Delta =\dfrac{d}{2}\pm \sqrt{\Big(\dfrac{d}{2}\Big)^2+
m_{\eta}^2} , \quad \quad
m_{\eta}^2=\dfrac{\ell(\ell+2)-8}{4},
\end{eqnarray*} 
one has 
\begin{eqnarray}
\Delta =1-\dfrac{\ell}{2}=1+\dfrac{k}{2}+q\bf{Z}. \label{ConB7}
\end{eqnarray}
From the superpotential in section 2, all $A, B, Q$ and $\tilde{Q}$ have the same conformal dimension $1/2$, 
thus dual mesonic operators of $\Delta$ are like   \begin{eqnarray}\label{tileQ}
\tilde{Q}^{I-1} A^{I}\big( A^I B^I \big)^{x_I}  \cdots 
A^{J} \big( A^J B^J \big)^{x_J} Q^{J+1},
~~~~~~\sum_K x_K = q {\bf Z}. 
\end{eqnarray}

As a remark, it is seen that meson spectra are not effected 
by gauge group ranks on different intervals. 
This can be seen from \eqref{D6action} where 
D6-brane DBI action has zero pull-back of NSNS 2-form flux $\propto \omega_2$ (K\"ahler form) arising from fractional M2-branes \cite{ABJ}.

\end{document}